\RequirePackage{fix-cm}
\documentclass[smallextended]{svjour3}       % onecolumn (second format)
\smartqed  % flush right qed marks, e.g. at end of proof

\usepackage{graphicx}
\usepackage{amssymb}
\usepackage{bm}
\usepackage{color}
\journalname{Journal of Low Temperature Physics}
\begin{document}

\title{Composite boson description of a low density gas of excitons}
% Grants or other notes about the article that should go on the front page should be placed here. General acknowledgments should be placed at the end of the article.
%\thanks{
%We acknowledge partial financial support from the MICINN (Spain) Grant No.~FIS2014-56257-C2-1-P.
%Yu. E. Lozovik was supported by RFBR.
%The Barcelona Supercomputing Center (The Spanish National Supercomputing Center -- Centro Nacional de Supercomputaci\'on) is %acknowledged for the provided computational facilities.
%The authors gratefully acknowledge the Gauss Centre for Supercomputing e.V. (www.gauss-centre.eu) for funding this project by %providing computing time on the GCS Supercomputer SuperMUC at Leibniz Supercomputing Centre (LRZ, www.lrz.de).
%}

\titlerunning{Composite boson description of a low density gas of excitons}
  % if too long for running head

\author{
A. E. Golomedov \and Yu. E. Lozovik    \\
\and G. E. Astrakharchik \and J. Boronat %etc.
}

\institute{
A. E. Golomedov\at
Yandex,\\
ulitsa Lva Tolstogo 16,\\
Moscow, Russia\\
119021
\and
Yu. E. Lozovik\at
Institute of spectroscopy RAS,\\
%Fizicheskaya Str., 5 \\
Troitsk, Moscow, Russia\\
108840
\and
G. E. Astrakharchik and J. Boronat\at
Departament de F\'\i sica, Universitat Polit\`ecnica de Catalunya \\
Barcelona, Spain\\
E-08034
}

\date{July 26, 2017}

\maketitle

\begin{abstract}

Ground state properties of a fermionic Coulomb gas are calculated using the fixed-node diffusion Monte Carlo method.
The validity of the composite boson description is tested for different densities.
We extract the exciton-exciton $s$-wave scattering length by solving the four-body problem in a harmonic trap and mapping the energy to that of two trapped bosons.
The equation of state is consistent with the Bogoliubov theory for composite bosons interacting with the obtained $s$-wave scattering length.
The perturbative expansion at low density has contributions physically coming from (a) exciton binding energy, (b) mean-field Gross-Pitaevskii interaction between excitons, (c) quantum depletion of the excitonic condensate (Lee-Huang-Yang terms for composite bosons).
In addition, for low densities we find a good agreement with the Bogoliubov bosonic theory for the condensate fraction of excitons.
The equation of state in the opposite limit of large density is found to be well described by the perturbative theory including (a) mixture of two ideal Fermi gases (b) exchange energy.
We find that for low densities both energetic and coherent properties are correctly described by the picture of composite bosons (excitons).
\end{abstract}

\section{Introduction\label{sec:introduction}}

The achievement of Bose-Einstein condensation (BEC) in confined alkali gases at nanokelvin temperatures has reinforced the interest in the search for other systems showing this extreme quantum behavior.
In this line, the progress achieved in recent years towards the observation of a BEC state in Coulomb systems based on electrons and holes in semiconductors is of particular interest.
This new candidate for a Bose condensate and superfluid state will show its macroscopic quantum behavior at much larger temperatures than BEC states in ultracold gases due to the much lower mass of the electron with respect to alkali atoms.
This feature, and its expected sufficiently large lifetime, makes the study of BEC in Coulomb systems extremely interesting.

Thinking on a BEC state, where the constituents are electrons and holes, leads immediately to the idea of formation of composite bosons where one electron and one hole, both of Fermi statistics, bind together.
This composite particle is termed {\em exciton} and is on the basis of the search for a BEC state in electronic matter.
Direct excitons are the ones in which electron and hole are not physically separated by any external potential, whereas indirect ones are carried out by physically separating electron and holes in two different layers with almost zero transition probability between them.
Indirect excitons are the most studied ones and constitute the most probable scenario for observing their Bose-Einstein condensation with the advantage of a substantially larger lifetime with respect to the direct ones.
In addition, spatially separated and coupled electrons and holes after Bose condensation can form nondissipating electric currents in separated layers due to superfluidity of condensed excitons\cite{LozovikYudson76,LozovikNishanov76,LozovikYudson78,LozovikBerman1996,LozovikBerman1997}.

The case of direct excitons has been less studied, probably in part due to the experimental difficulty of making the system stable for a finite lifetime.
However, a gas of excitons is a clean and very interesting system from the theoretical side.
It is particularly interesting to study its properties in the limit of low densities in which a description of the system in terms of composite bosons might be appropriate.
Considering a gas of polarized electrons (treated as spin up particles) and polarized holes (spin down particles), the ground state at low density will be constituted by a gas of excitons where one electron and one hole couple and form a boson with integer spin.
Then, these composite bosons will behave as bosons with a mass equal to the sum of the masses of electron and hole and the $s$-wave scattering length between excitons will be the dominant parameter of their effective interaction.
In some sense, this is formally equivalent to the formation of molecules in dilute two-component Fermi gases with positive
scattering lengths, i.e., beyond the unitary limit.

While the excitonic description is very simple and tempting, due to the possibility of using well-established techniques (Gross-Pitaevskii equation\cite{Gross61,Pitaevskii61}, Bogoliubov theory\cite{Lifshitz80}, etc), in the last years there was a strong criticism of the very idea of such possibility.
One of the strongest opponents to such description comes notably due to Monique Combescot who by introducing Shiva diagrams and performing calculations\cite{CombescotBook} argued that the composite-boson description of an exciton intrinsically misses a relevant part.
That is, for some properties an elementary boson differs in a fundamental way from two Coulomb fermions due to the composite nature which prohibits\cite{PhysRevB.75.174305} to describe the interaction between excitons by some effective potential even in the extremely low density limit, and eventually to make use of the usual many-body theories.
Also in a classical work\cite{KeldyshKozlov68} by Keldysh and Kozlov dating back to 1968, it was shown that the commutation relation for the exciton creation $\hat a_{\bf k}$ and annihilation $\hat a^\dagger_{\bf k}$ operators do not obey usual Bose commutation rule as, instead, $[\hat a_{\bf k},\hat a_{\bf k'}] = \delta_{{\bf k},{\bf k'}} + O(na_0^3)$,
where $n$ is the particle density and $a_0$ the Bohr radius.
As a result there, the contribution due non-bosonic commutation relation is of the order of $na_0^3$ and consequently it induces corrections to the energy already in the lowest order of the density, that is exactly on the same level as coming from the mean-field theory for Bose particles.

It is interesting to check if the bosonic or fermionic nature of excitons manifests itself in the energetic and coherent properties of the gas.
If the composite-boson description is possible, the equation of state can be expanded in powers of the gas parameter $na^3$, where $a$ is the $s$-wave scattering length coming from the four-body problem (note that $a$ effectively includes exchange effects).
As a function of density, the mean-field contribution to the energy per particle should scale as $\propto n$ and beyond-mean field one as $\propto n^{3/2}$.
In addition, the equation of state might contain terms proportional to the Fermi momentum $k_F = (3\pi^2 n)^{1/3} \propto n^{1/3}$ or the Fermi energy $\propto n^{2/3}$.
Thus, by calculating the expansion of the equation of state in an {\em ab initio} microscopic simulation of a Coulomb fermionic system we should be able to see which description holds.
Furthermore, we can check if the exciton-exciton interaction can be described in terms of some effective potential, namely a short-range potential with an effective $s$-wave scattering length $a_{\rm{ee}}$.
To do so we can first solve the four-body problem and extract $a_{\rm{ee}}$ and afterwards compare the energy in the many-body system.

It can be argued, that Quantum Monte Carlo methods are extremely well suited for studying the equilibrium properties of electron and Coulomb systems.
Fixed-node diffusion Monte Carlo calculations of jellium surfaces were performed by Acioli and Ceperley\cite{AcioliCeperley96}.
A relativistic electron gas was studied by VMC and DMC methods by Kenny et al.\cite{PhysRevLett.77.1099}
The electron-hole plasma was recently studied by variational Monte Carlo\cite{Zhu96}
and diffusion Monte Carlo\cite{Senatore2002,DrummondNeeds16} approaches.
Two-dimensional electron gas in strong magnetic fields was investigated in Ref.~\cite{PhysRevB.54.4948} by means of the variational Monte Carlo method.
Finite-temperature properties can be accessed using path integral Monte Carlo method.
The high-temperature phase diagram of a hydrogen plasma was obtained in Ref.~\cite{PhysRevLett.76.1240}.
The biexciton wave function was obtained using a quantum Monte Carlo calculation in Ref.~\cite{Littlewood13}.

In the present paper, we analyze a gas of excitons at low densities trying to verify if their description as composite bosons is compatible with the low density expansion for the energy and condensate fraction of a dilute universal Bose gas.
To this end, we have performed quantum Monte Carlo simulations of the Fermi electron-hole gas using accurate trial wave functions and the fixed-node approximation to control the sign.
To make the comparison feasible we have also calculated the scattering length between excitons which shows agreement with
previous estimations.
At low densities, the effective description of energy and condensate fraction of pairs is fully compatible with the universal law for dilute bosons without any significant contribution of purely Fermi contributions.

The rest of the paper is organized as follows.
In Section~\ref{Sec:QMC}, we briefly describe the quantum Monte Carlo method used in the present study.
Section~\ref{Sec:4body} comprises the analysis of the four-body problem in harmonic confinement used
to determine the exciton-exciton $s$-wave scattering length.
Results of the many-body problem and their effective description as composite bosons are reported in Sec.~\ref{Sec:eh gas}.
Finally, we draw the conclusions of the work in Sec.~\ref{Sec:conclusions}.

\section{Quantum Monte Carlo method\label{Sec:QMC}}

In the present work the electron-hole system is microscopically described using the diffusion Monte Carlo (DMC) method.
DMC is nowadays a standard tool for describing quantum many-body systems that solves, in a stochastic way, the imaginary-time
Schr\"odinger equation of the system (for a general reference on the DMC method, see for example~\cite{BoronatCasulleras94}).
For particles obeying Bose-Einstein statistics, DMC solves exactly the problem for the ground state within some statistical variance.
When the system under study is of Fermi type we need to introduce an approximation to account for the non-positiveness of the wave function.
This approximation, known as {\em fixed node} (FN), restricts the random walks within the nodal pockets defined by a trial wave function used as importance sampling technique during the imaginary-time evolution.
Further details on the FN-DMC method can be found elsewhere.

Our system is composed by a mixture of $N_e$ electrons with mass $m_e$ and $N_h$ holes with mass $m_h$.
All the electrons (holes) have the same spin up (down).
The Hamiltonian of the system is
\begin{equation}
H=-\frac{\hbar^2}{2 m_e} \sum_{i=1}^{N_e} {\bm \nabla}_i^2
-\frac{\hbar^2}{2 m_h} \sum_{i^\prime=1}^{N_h} {\bm \nabla}_{i^\prime}^2
+\sum_{i<j}^{N_e} \frac{e^2}{r_{ij}} +\sum_{i^\prime<j^\prime}^{N_h}
\frac{e^2}{r_{i^\prime j^\prime}}
-\sum_{i,i^\prime=1}^{N_e,N_h} \frac{e^2}{r_{ii^\prime}} \ ,
\label{hamiltonian}
\end{equation}
where $i,j,\ldots$ and $i^\prime,j^\prime,\ldots$ label electron and hole coordinates, respectively.
In our study, we have considered equal masses $m_e=m_h\equiv m$ and used distances measured in units of the Bohr radius $a_0=\hbar^2/(m e^2)$ and energies in Hartrees, $1\,{\rm Ha}=e^2/a_0$.
Therefore, in these units the Hamiltonian becomes
\begin{equation}
H=-\frac{1}{2} \sum_{i=1}^{N_e} {\bm \nabla}_i^2
-\frac{1}{2} \sum_{i^\prime=1}^{N_h} {\bm \nabla}_{i^\prime}^2
+\sum_{i<j}^{N_e} \frac{1}{r_{ij}} +\sum_{i^\prime<j^\prime}^{N_h}
\frac{1}{r_{i^\prime j^\prime}}
-\sum_{i,i^\prime=1}^{N_e,N_h} \frac{1}{r_{ii^\prime}} \ .
\label{hamiltonianr}
\end{equation}

The convergence of DMC method can be significantly improved by a proper choice of the trial wave function used for the importance sampling.
As we are interested in the excitonic phase at low densities, our model for the wave function in the superfluid phase is
\begin{equation}
\Psi(\bm{R}) = {\cal A} (\phi(r_{1 1^\prime})  \phi(r_{2 2^\prime})
\ldots \phi(r_{N_e N_h})) \ ,
\label{trial}
\end{equation}
with ${\cal A}$ the antisymmetrizer operator of all the pair orbitals $\phi(r_{i i\prime})$.
This function is taken from the ground-state solution of the two-body problem, $\phi(r_{i i^\prime})= \exp[-r_{i i^\prime}/(2 a_0)]$, corresponding to the electron-hole bound state with energy $E_b=-\hbar^2/(4 m a_0^2)$.
It is worth noticing that a similar approach\cite{DMCBECBCS,AstrakharchikGiorginiBoronat2012} was used in the study of the unitary limit of a two-component Fermi gas and proved its accuracy in reproducing the experimental data.

In order to take into account the long-range behavior of the Coulomb interaction, we used standard Ewald summation to reduce size effects.
Other possible bias coming from the use of a finite time step and number of walkers were optimized to reduce their
effect to the level of the typical statistical noise.

\section{Four body problem. Exciton-exciton scattering length \label{Sec:4body}}

If the description of excitons in terms of composite bosons is possible, the size of each composite boson is of the order of the Bohr radius $a_0$ which becomes small in the limit of dilute density, $n a_0^3\to 0$.
The induced-dipole interaction between electron-hole pairs is of a Van der Waals type with $1/r^6$ decay at large distances and can be treated as a short-range potential in that limit.
This suggests that in the limit of dilute density, the exciton-exciton interaction potential can be described by a single parameter, the $s$-wave scattering length $a_{\rm ee}$.
In this section we extract its value from the four-body problem.
A textbook procedure\cite{LandauLifshitz_volume_III} of finding the $s$-wave scattering length involves finding the low-energy asymptotic of the phase shift in the scattering problem.
Alternatively, one might solve the few-body problem in a harmonic oscillator trapping and map the energy to that of a two-boson problem and take the limit of the vanishing strength of the trap\cite{KanjilalBlume2008,Blume2012}

We calculate the energy of the 1e+1h and 2e+2h systems confined in a harmonic trap of different frequencies.
The Hamiltonian in this case is the sum of the original Hamiltonian $H$, Eq.~(\ref{hamiltonianr}), and the confining term, that is
\begin{equation}
H_c = H + \sum_{i=1}^{N_e} \frac{1}{2}r_i^2 + \sum_{i^\prime=1}^{N_h} \frac{1}{2}r_{i^\prime}^2 \ ,
\label{hamiltonianc}
\end{equation}
where we consider equal masses $m_e=m_h=m$ and use harmonic oscillator (HO) dimensionless units, that is HO length $a_{\rm ho}=\sqrt{\hbar/(m \omega)}$ for distances and HO level spacing $E_0=\hbar \omega$ for the energies.
To improve the sampling, the trial wave function~(\ref{trial}) is multiplied by one-body terms which are the solution of non-interacting particles under the harmonic confinement,
\begin{equation}
\Psi_c(\bm{R}) = \prod_{i=1}^{N_e} e^{-\alpha r_i^2}\prod_{i^\prime=1}^{N_h} e^{-\alpha r_{i^\prime}^2} \, \Psi(\bm{R}) \ .
\label{trialc}
\end{equation}

The two-body problem, 1e-1h, can be solved exactly using a numerical grid method and also using the DMC method.
We have verified that both results match exactly.
For the four-body case, 2e-2h, we deal only with the DMC method.
The energies for the two and four-body problems can be split in the following form
\begin{eqnarray}
E_2 & = & E_b + E_{\rm CM} \\
E_4 & = & 2 E_b + E_{\rm int} + E_{\rm CM} \ ,
\label{enersplit}
\end{eqnarray}
with $E_b$ the binding energy of 1e-1h, $E_{\rm CM}=3/2$ the center-of-mass energy, and $E_{\rm int}$ the energy associated to the exciton-exciton interaction.
We are mainly interested in the last one,
\begin{equation}
E_{\rm int} = (E_4-2 E_2) + \frac{3}{2} \ ,
\label{enerexciton}
\end{equation}
because from it we can extract the $s$-wave scattering length.

The 2e-2h system in a harmonic trap can be thought as forming two dimers (excitons) obeying Bose statistics.
These composite bosons interact with some short-range potential, which can be approximated as a regularized contact pseudopotential,
\begin{equation}
V(r) = 4 \pi a_{\rm ee} \delta({ \mathbf r}) \frac{\partial}{\partial r} (r\cdot)  \ .
\label{pseudo_potential}
\end{equation}
Within this approximation, one ends up with a problem of two bosons in a harmonic trap described by effective Hamiltonian
\begin{equation}
H_2^b = -\frac{1}{2} \nabla_{1,2}^2 +\frac{1}{2} r_{1,2}^2 + 4 \pi a_{\rm ee} \delta({\mathbf r}_{12}) \frac{\partial}{\partial r_{12}} (r_{12}\cdot)  \ ,
\label{hamiltonian_composite_bosons}
\end{equation}
with $m_b$ being the mass of composite particle ($m_b = m_e+m_h = 2 m$ in the case of equal masses).
The eigenstates of Hamiltonian~(\ref{hamiltonian_composite_bosons}) can be found analytically, see Ref~\cite{Busch98}), and the $s$-wave scattering length $a_{\rm ee}$ can be found as a solution of the following equation, 
\begin{equation}
a_{\rm ee} = \frac{1}{\sqrt{2}} \, \frac{\Gamma(-E_{\rm int}/2+1/4)}{\Gamma(-E_{\rm int}/2+3/4)} \ ,
\label{eqn:Energy_a_dd}
\end{equation}
with $E_{\rm int}$ the energy associated to the exciton-exciton interaction~(\ref{enerexciton}).

\begin{figure}
\begin{center}
\includegraphics[width=0.8\linewidth]{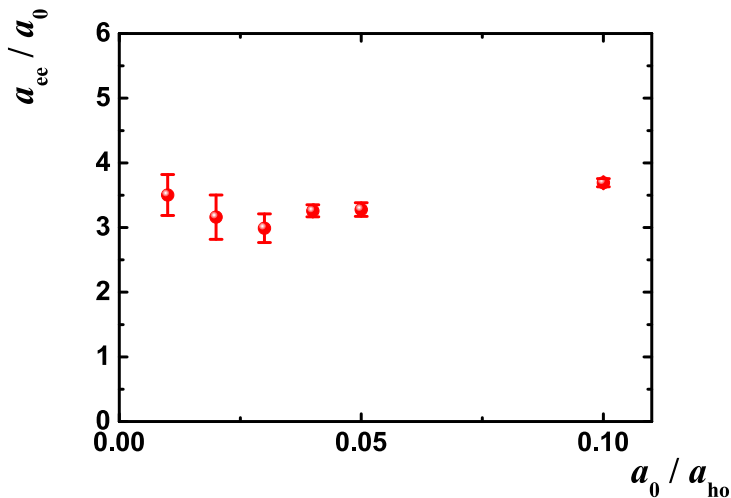}
\end{center}
\caption{Exciton-exciton $s$-wave scattering length as a function of $a_0/a_{\rm ho}$. }
\label{fig:scatering}
\end{figure}

Results for the scattering length $a_{\rm ee}$ obtained through the combination of DMC results for the energy $E_{\rm int}$ and the formula for $a_{\rm ee}$ (\ref{eqn:Energy_a_dd}) are reported in Fig.~\ref{fig:scatering} for different values of $a_0/a_{\rm ho}$.
As one can see, the dependence of  $a_{\rm ee}$ on the strength of the confinement is rather shallow, approaching a value
$a_{\rm ee}\simeq 3 a_{0}$ when $a_0/a_{\rm ho} \to 0$.
This result is in nice agreement with previous estimations by Shumway and Ceperley based on finite-temperature calculations performed using the path integral Monte Carlo method\cite{Shumway2000} and DMC phase-shift calculations\cite{ShumwayCeperley2001,ShumwayCeperley2005}. 
We will use this value for making a comparison with the Bogoliubov theory for the many-body problem.

\section{Electron-hole gas\label{Sec:eh gas}}

Using the formalism introduced in Sec.~\ref{Sec:QMC} we have calculated the properties of a bulk electron-hole gas, mainly for very low values of the gas parameter $na_0^3$, with $n=(N_e+N_h)/V$ the total density.
We consider an unpolarized gas, $N_e = N_h$, of equal mass particles.
As we are interested in the description of the excitonic phase we use as a trial wave function a determinant composed by electron-hole orbitals (see Sec.~\ref{Sec:QMC}).

\begin{figure}
\begin{center}
\includegraphics[width=0.8\linewidth]{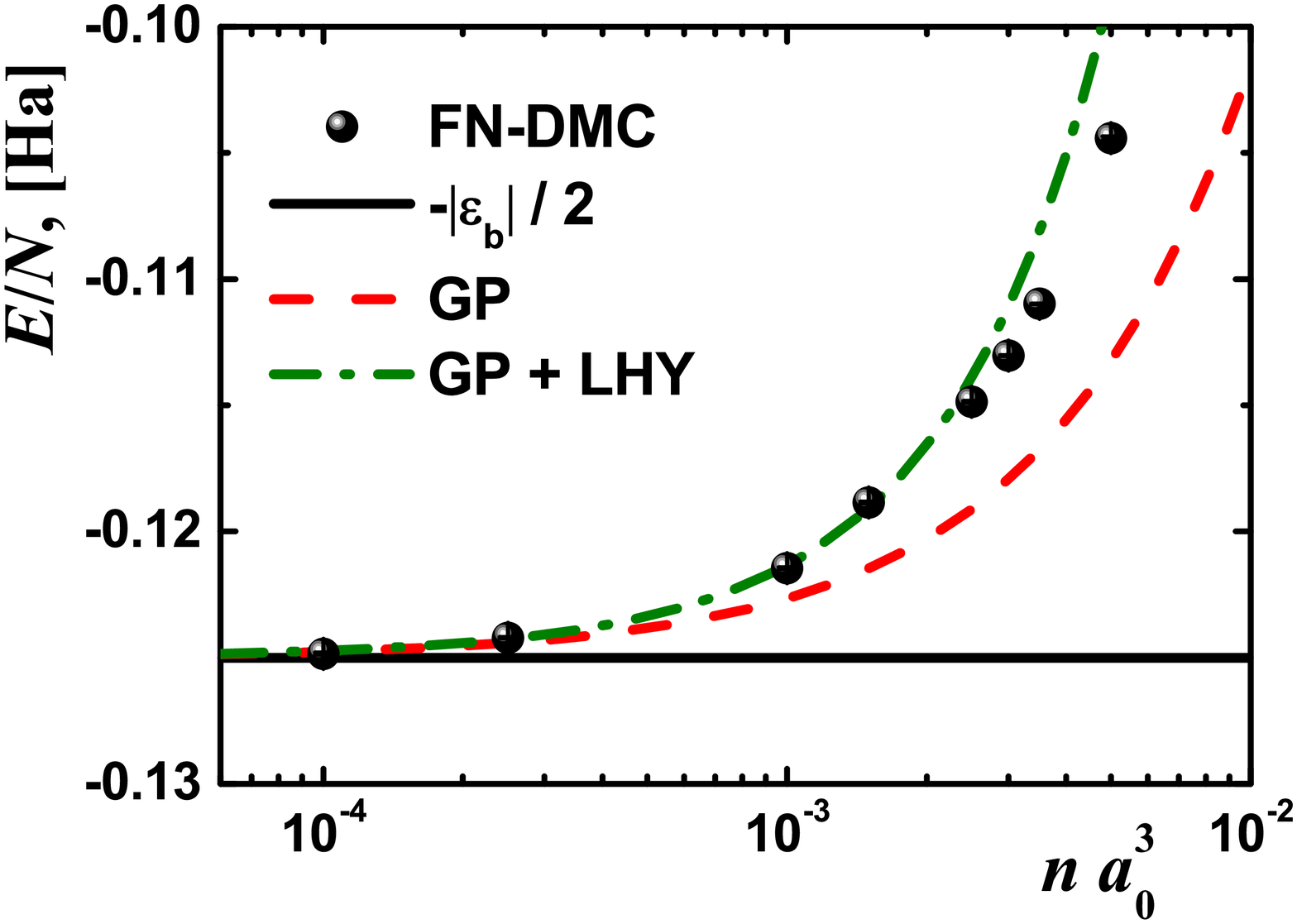}
\end{center}
\caption{FN-DMC energies per particle as a function of the gas parameter $na_0^3$.
In the limit of zero density we recover the binding energy of the electron-hole pair.
The dashed and dot-dashed lines stand for the MF and MF+LHY energies, respectively,  considering the exciton-exciton scattering length $a_{\rm{ee}}=3 a_0$.
The lines are shifted to give the binding energy of the pair at zero density.
}
\label{fig:lowenergy}
\end{figure}

In Fig.~\ref{fig:lowenergy}, we plot the energy of the electron-hole gas per particle $E/N$ as a function of the gas parameter $na_0^3$.
At very low densities, $na_0^3 \lesssim 10^{-4}$, the energy per particle tends to half the binding energy of an electron-hole pair, $-|\varepsilon_b|/2=-0.125$ Ha.
When the density increases the energy also increases due to the repulsive interaction between excitons.
Within the picture of composite bosons, the mean-field Gross-Pitaevskii energy\cite{Lifshitz80,book_pitaevskii.stringari.2016} of a weakly interacting composite Bose gas can be written as
\begin{equation}
\left(\frac{E}{N_{ex}} \right)_{\rm{MF}} =
\frac{1}{2} g_{ex} n_{ex}\;,
\label{mfield:GP}
\end{equation}
where number of excitons is twice smaller than the number of charges, $N_{ex} = N/2$,
and corresponding concentration is twice smaller, $n_{ex} = n/2$,
exciton has a twice larger mass, $m_{ex} = 2 m$,
the coupling constant between excitons is $g_{ex} = 4\pi \hbar^2 a_{\rm ee}/ m_{ex}$
with $a_{\rm ee}$ being the exciton-exciton $s$-wave scattering length.
Equation~\ref{mfield:GP} can be written in atomic Hartee units as
\begin{equation}
\left( \frac{E}{N} \right)_{\rm{MF}} = \frac{\pi}{4a_{\rm{ee}}^2} \, n a_{\rm{ee}}^3\;.
\label{mfield}
\end{equation}
In Fig.~\ref{fig:lowenergy}, we plot the mean-field energy~(\ref{mfield}) shifted to be half the binding energy of the pair $-|\epsilon_b|/2$ and compare it with the results of FN-DMC calculations.
Our results match the mean-field energy with $a_{\rm{ee}}=3 a_0$ at very low densities, $na_0^3 \lesssim 10^{-4}$ but, when the gas parameter increases more, the FN-DMC energies increase faster than the mean-field law.
Adding the Lee-Huang-Yang (LHY) correction\cite{Huang57,Lee57} to the mean-field term~(\ref{mfield}),
\begin{equation}
\left( \frac{E}{N} \right)_{\rm{LHY}}
= \frac{\pi}{4 a_{\rm{ee}}^2} \, n a_{\rm{ee}}^3 \left[ 1 + \frac{128}{15
\sqrt{\pi}} \sqrt{n_{\rm{ex}} a_{\rm{ee}}^3}\right]
\label{lhy}
\end{equation}
we can estimate the beyond-mean-field first correction.
In Fig.~\ref{fig:lowenergy}, we plot LHY energy~(\ref{lhy}) to be compared with the DMC data.
As one can see, the LHY law reproduces our data up to densities $na_0^3 \sim 3\cdot 10^{-3}$ which approach the end of the universal regime, where the energy of a Bose gas is completely described solely in terms of the
gas parameter.
The LHY term arises from quantum fluctuations of the bosons that drop out of the condensate and in a single component LHY correction is accurate up to $na^3 \lesssim 10^{-3}$\cite{Giorgini99}, where $a$ is the boson-boson $s$-wave scattering length. It is interesting to note that the energetic behavior of the Coulomb electron-hole gas at low-density is fully described by the picture of composite bosons.
These results corroborate the picture of an exciton as being considered effectively as a composite boson.

\begin{figure}
\begin{center}
\includegraphics[width=0.8\linewidth]{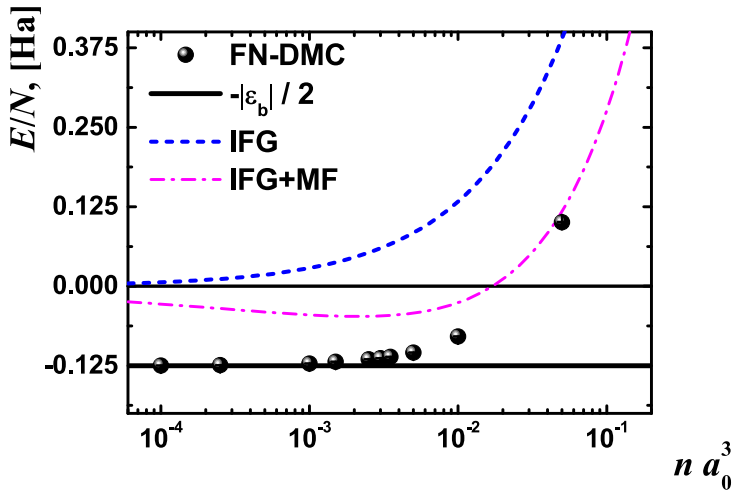}
\end{center}
\caption{FN-DMC energies per particle as a function of the gas parameter $na_0^3$.
The dashed line stands for the energy of a mixture of two ideal Fermi gases.
The dotted-dashed lines incorporates the exchange energy to the ideal Fermi gas model. }
\label{fig:highenergy}
\end{figure}

When the density increases even more the energies depart from the low-density universal expansion~(\ref{lhy}).
At high density one expects that the system evolves to a mixture of two ideal Fermi gases with energy\cite{Lifshitz80}
\begin{equation}
\frac{E^{(0)}}{N}= \frac{3}{10} \frac{\hbar^2k_F^2}{m}\ ,
%\label{fermigas}
\end{equation}
with the Fermi wave number $k_F = (3\pi^2 n)^{1/3}$ or, the energy per exciton as a function of Wigner-Seitz radius $r_s = [3/(4 \pi n)]^{1/3} / a_0$ in Hartree atomic units,
\begin{equation}
\frac{E^{(0)}}{N_{\rm{ex}}}= \frac{2.21}{r_s^2} \ .
\label{fermigas}
\end{equation}
In Fig.~\ref{fig:highenergy}, we plot the energy~(\ref{fermigas}) as a function of the gas parameter $na_0^3$.
As one can see this energy is clearly out of our results.
However, if one incorporates the exchange energy derived as a first-order perturbation theory on top of the free Fermi gas\cite{FetterWalecka,GiulianiVignaleBook},
\begin{equation}
% Eq. (1.107) Gabriele F. Giuliani and Giovanni Vignale
\frac{E^{(1)}}{N_{\rm{ex}}}= \frac{2.21}{r_s^2} - \frac{0.916}{r_s} \ .
\label{exchange}
\end{equation}
our results approach well to Eq.~(\ref{exchange}).

\begin{figure}
\begin{center}
\includegraphics[width=0.8\linewidth]{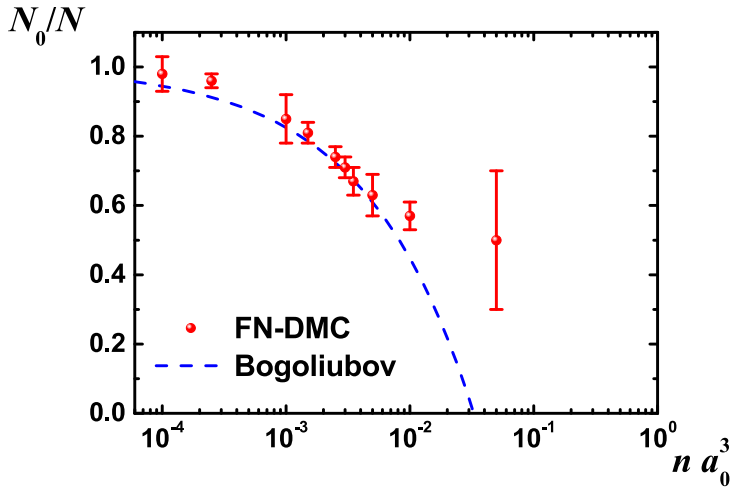}
\end{center}
\caption{Condensate fraction of excitons as a function of the gas parameter
$na_0^3$. The line corresponds to the Bogoliubov prediction for a dilute gas of
composite bosons interacting with $s$-wave scattering length $a_{\rm{ee}}$.   }
\label{fig:condensate}
\end{figure}

If the description of excitons as composite bosons, interacting with an effective $s$-wave scattering length $a_{\rm{ee}}$, is  correct at low densities then we have to observe a finite fraction of condensate pairs.
We found that the excitonic picture of composite bosons provides a good energetic description and it is important to verify up to which level the excitonic description is valid in terms of the coherence in the correlation functions.
To this end, we have calculated the two-body density matrix
\begin{equation}
\rho_2({\bf r}_1^\prime,{\bf r}_2^\prime,{\bf r}_1,{\bf r}_2)=
\langle\psi_\uparrow^\dagger({\bf r}_1^\prime)
\psi_\downarrow^\dagger({\bf r}_2^\prime)\psi_\uparrow({\bf r}_1)
\psi_\downarrow({\bf r}_2)\rangle \;.
\label{TBDM}
\end{equation}
For an unpolarized gas with $N_e=N_\uparrow=N/2$ and $N_h=N_\downarrow=N/2$, if $\rho_2$ has an eigenvalue of the order of the total number of particles $N$, the $\rho_2$ can be
decomposed as,
\begin{equation}
\rho_2({\bf r}_1^\prime,{\bf r}_2^\prime,{\bf r}_1,{\bf r}_2)=
\alpha N/2 \varphi^\ast({\bf r}_1^\prime,{\bf r}_2^\prime)
\varphi({\bf r}_1,{\bf r}_2) + \rho_2^\prime \;,
\label{TBDM1}
\end{equation}
$\rho_2^\prime$ containing only eigenvalues of order one.
The parameter $\alpha\le 1$ in Eq.~(\ref{TBDM1}) is interpreted as the condensate fraction of pairs (excitons), in a similar way as the condensate fraction of single atoms is derived from the one-body density matrix.

The spectral decomposition (\ref{TBDM1}) yields for homogeneous systems the following asymptotic  behavior of $\rho_2$
\begin{equation}
\rho_2({\bf r}_1^\prime,{\bf r}_2^\prime,{\bf r}_1,{\bf r}_2) \to \alpha N/2
\varphi^\ast(|{\bf r}_1^\prime-{\bf r}_2^\prime|)
\varphi(|{\bf r}_1-{\bf r}_2|) \;,
\label{TBDM2}
\end{equation}
if $|{\bf r}_1-{\bf r}_1^\prime|$, $|{\bf r}_2-{\bf r}_2^\prime|\to\infty$.
The wave function $\varphi$ is proportional to the order parameter $\langle\psi_\uparrow({\bf r}_1)\psi_\downarrow({\bf r}_2)\rangle=\sqrt{\alpha N/2} \varphi(|{\bf r}_1-{\bf r}_2|)$, whose appearance characterizes the superfluid state of composite bosons.

In Fig.~\ref{fig:condensate}, we plot the condensate fraction of excitons as a
function of the gas parameter. At very low densities practically all the pairs
are in the condensate, $N_0/N \to 1$ and this value decreases monotonically
towards zero with the density. The DMC estimation of the condensate fraction
becomes difficult at large densities, which translates into a larger
statistical noise, as can be appreciated in the figure. When the gas parameter
is low enough one expects to recover the Bogoliubov law,
\begin{equation}
\frac{N_0}{N}= 1 - \frac{8}{3 \sqrt{\pi}}  \sqrt{n_{\rm{ex}} a_{\rm{ee}}^3} \ .
\label{condens}
\end{equation}
We compare this low density universal behavior~(\ref{condens}) with the DMC data in Fig.~\ref{fig:condensate}.
As we can see, the agreement is excellent corroborating that the composite-boson picture with $a_{\rm{ee}}$ is fully consistent. It is interesting to note that the the universal behavior in a single component Bose gas breaks down at a similar value of the gas parameter, $na^3 \sim 10^{-2}$\cite{Giorgini99}.

\section{Conclusions\label{Sec:conclusions}}

The consideration of excitons as composite bosons has been controversial for many years.
Our DMC calculations have tried to contribute to this discussion using a microscopic approach, with the only restriction of the fixed-node approximation to overcome the sign problem.
Working first with a four-body problem we have obtained the $s$-wave scattering length of the exciton-exciton interaction.
The value obtained is in good agreement with previous estimations obtained in finite-temperature path integral Monte Carlo calculations.
In the second part of the present study, we have calculated the properties of a homogeneous electron-hole system, focusing on the energy and the excitonic condensate fraction.
Both the energy and condensate fraction agrees perfectly at low densities with the universal relations in terms of the gas parameter.
Using the scattering length, obtained from the four-body problem, we reproduce the DMC data at low densities with good accuracy. In particular, we observe the relevance of the Lee-Huang-Yang term, beyond the mean field one, in describing correctly
the energy.
It is important to note that exchange terms appearing due to composite structure of excitons, as build up from excitons, is taken into account in the way we solve the four-body problem.
As a result, the exchange effects are readily incorporated in the effective exciton-exciton $s$-wave scattering length $a_{\rm ee}=3a_0$.
Only after the universal regime breaks down, the energies depart from the composite-boson picture and approach the regime of a Fermi gas with Coulomb interaction.
The equation of state in the high-density regime agrees with the description in terms of the energy of two ideal Fermi gases corrected by the exchange energy arising due to Coulomb interactions.
With respect to the condensate fraction of excitons, we have verified by means of a calculation of the two-body density matrix that the condensate fraction of pairs matches the Bogoliubov prediction of a Bose gas of particles interacting with an scattering length $a_{\rm{ee}}$ for low values of the gas parameter.

Altogether, our results allow to conclude that the disputed interpretation of excitons as composite bosons is actually consistent in terms of energy and coherence with our results once the effective $s$-wave scattering length is extracted from the four-body problem using energy mapping to a two-boson problem.

\begin{acknowledgements}

Pierbiagio Pieri and Alexander Fetter are acknowledged for useful discussions about the expansion of the equation of state for a weakly-interacting Fermi gas.
We acknowledge partial financial support from the MICINN (Spain) Grant No.~FIS2014-56257-C2-1-P.
Yu. E. Lozovik was supported by RFBR.
The authors thankfully acknowledge the computer resources at MareNostrum and the technical support provided by Barcelona Supercomputing Center (FI-2017-2-0011).
The authors gratefully acknowledge the Gauss Centre for Supercomputing e.V. (www.gauss-centre.eu) for funding this project by providing computing time on the GCS Supercomputer SuperMUC at Leibniz Supercomputing Centre (LRZ, www.lrz.de).

\end{acknowledgements}

%% BibTeX users please use one of
%%\bibliographystyle{spbasic}      % basic style, author-year citations
%%\bibliographystyle{spmpsci}     % mathematics and physical sciences
%\bibliographystyle{spphys}      % APS-like style for physics

%\bibliography{excitonqmc}

\end{document}